\documentclass{ws-ijmpb}
\usepackage{epsfig}
\usepackage{graphicx}
\chardef\quoteleftcode=\catcode96	
\newread\epsffilein    
\newif\ifepsffileok    
\newif\ifepsfbbfound   
\newif\ifepsfverbose   
\newdimen\epsfxsize    
\newdimen\epsfysize    
\newdimen\epsftsize    
\newdimen\epsfrsize    
\newdimen\epsftmp      
\newdimen\pspoints     
\def\epsfbox#1{\global\def\epsfllx{72}\global\def\epsflly{72}%
   \global\def\epsfurx{540}\global\def\epsfury{720}%
   \def\lbracket{[}\def\testit{#1}\ifx\testit\lbracket
   \let\next=\epsfgetlitbb\else\let\next=\epsfnormal\fi\next{#1}}%
\def\epsfgetlitbb#1#2 #3 #4 #5]#6{\epsfgrab #2 #3 #4 #5 .\\%
   \epsfsetgraph{#6}}%
\def\epsfnormal#1{\epsfgetbb{#1}\epsfsetgraph{#1}}%
\def\epsfgetbb#1{%
\openin\epsffilein=#1
\ifeof\epsffilein\errmessage{I couldn't open #1, will ignore it}\else
   {\epsffileoktrue \chardef\other=12
    \def\do##1{\catcode`##1=\other}\dospecials \catcode`\ =10
    \catcode`\^^L=9 \catcode`\^^?=9
    \loop
       \read\epsffilein to \epsffileline
       \ifeof\epsffilein\epsffileokfalse\else
	  \expandafter\epsfaux\epsffileline:. \\%
       \fi
   \ifepsffileok\repeat
   \ifepsfbbfound\else
    \ifepsfverbose\message{No bounding box comment in #1; using defaults}\fi\fi
   }\closein\epsffilein\fi}%
\def\epsfclipstring{}
\def\epsfsetgraph#1{%
   \epsfrsize=\epsfury\pspoints
   \advance\epsfrsize by-\epsflly\pspoints
   \epsftsize=\epsfurx\pspoints
   \advance\epsftsize by-\epsfllx\pspoints
   \epsfxsize\epsfsize\epsftsize\epsfrsize
   \ifnum\epsfxsize=0 \ifnum\epsfysize=0
      \epsfxsize=\epsftsize \epsfysize=\epsfrsize
      \epsfrsize=0pt
     \else\epsftmp=\epsftsize \divide\epsftmp\epsfrsize
       \epsfxsize=\epsfysize \multiply\epsfxsize\epsftmp
       \multiply\epsftmp\epsfrsize \advance\epsftsize-\epsftmp
       \epsftmp=\epsfysize
       \loop \advance\epsftsize\epsftsize \divide\epsftmp 2
       \ifnum\epsftmp>0
          \ifnum\epsftsize<\epsfrsize\else
	     \advance\epsftsize-\epsfrsize \advance\epsfxsize\epsftmp \fi
       \repeat
       \epsfrsize=0pt
     \fi
   \else \ifnum\epsfysize=0
     \epsftmp=\epsfrsize \divide\epsftmp\epsftsize
     \epsfysize=\epsfxsize \multiply\epsfysize\epsftmp
     \multiply\epsftmp\epsftsize \advance\epsfrsize-\epsftmp
     \epsftmp=\epsfxsize
     \loop \advance\epsfrsize\epsfrsize \divide\epsftmp 2
     \ifnum\epsftmp>0
	\ifnum\epsfrsize<\epsftsize\else
	   \advance\epsfrsize-\epsftsize \advance\epsfysize\epsftmp \fi
     \repeat
     \epsfrsize=0pt
    \else
     \epsfrsize=\epsfysize
    \fi
   \fi
   \ifepsfverbose\message{#1: width=\the\epsfxsize, height=\the\epsfysize}\fi
   \epsftmp=10\epsfxsize \divide\epsftmp\pspoints
   \vbox to\epsfysize{\vfil\hbox to\epsfxsize{%
      \ifnum\epsfrsize=0\relax
        \includegraphics{#1}%
      \else
        \epsfrsize=10\epsfysize \divide\epsfrsize\pspoints
        \includegraphics{#1}%
      \fi
      \hfil}}%
\global\epsfxsize=0pt\global\epsfysize=0pt}%
{\catcode`\%=12 \global\let\epsfpercent=
\long\def\epsfaux#1#2:#3\\{\ifx#1\epsfpercent
   \def\testit{#2}\ifx\testit\epsfbblit
      \epsfgrab #3 . . . \\%
      \epsffileokfalse
      \global\epsfbbfoundtrue
   \fi\else\ifx#1\par\else\fi\fi}%
\def\epsfempty{}%
\def\epsfgrab #1 #2 #3 #4 #5\\{%
\global\def\epsfllx{#1}\ifx\epsfllx\epsfempty
      \epsfgrab #2 #3 #4 #5 .\\\else
   \global\def\epsflly{#2}%
   \global\def\epsfurx{#3}\global\def\epsfury{#4}\fi}%
\def\epsfsize#1#2{\epsfxsize}
\catcode`\`=\quoteleftcode		

\begin{document}

\markboth{G. Cordourier-Maruri, R. de Coss, and V. Gupta}
{Transmission properties of the one-dimensional array of delta potentials}

%
\catchline{}{}{}{}{}
%

\title{TRANSMISSION PREPERTIES OF THE ONE-DIMENSIONAL ARRAY OF DELTA POTENTIALS}

\author{GUILLERMO CORDOURIER-MARURI$^*$, ROMEO DE COSS$^{\dagger}$, AND VIRENDRA GUPTA$^{\ddagger}$}

\address{Departamento de F\'isica Aplicada, Centro de Investigaci\'on y Estudios Avanzados del IPN \\ 
Unidad M\'erida, A.P. 73 Cordemex, M\'erida, Yucat\'an, 97310, M\'exico\\
* gmaruri@mda.cinvestav.mx\\
$^{\dagger}$ decoss@mda.cinvestav.mx\\
$^{\ddagger}$ virendra@mda.cinvestav.mx}

\maketitle

\begin{history}
\received{Day Month Year}
\revised{Day Month Year}
\end{history}

\begin{abstract}
                  The problem of one-dimensional quantum wire along which a moving particle interacts with a linear array of
                  $N$ delta-function potentials is studied. Using a quantum waveguide approach, the transfer matrix is calculated to obtain
                  the transmission probability of the particle. Results for arbitrary $N$ and for specific regular arrays are presented. Some
                  particular symmetries and invariances of the delta-function potential array for the $N = 2$ case are analyzed in
                  detail. It is shown that perfect transmission can take place in a variety of situations.\end{abstract}


\section{Introduction}\label{intro}

     The use of one-dimensional potential arrays is very frequent in several fields of physics. In solid state physics, the one-dimensional         periodic arrays are used to model the lattice in a crystal, as a first approach. Several works use similar formulation to demonstrate
     Anderson localization in different types of lattice disorder \cite{and,hue,koh,si}. 
        
     A delta-function is useful to describe short-range potentials, like the interaction between the electrons and fixed ions in a lattice
     crystal. Thus, a periodic array of delta-function potentials is used in the Kronig-Penney model \cite{kp}. On the other hand, the 
     delta-potential is also useful to describe impurities in solid state systems. Thus, the study of electron scattering by impurities in
     quantum wires, using delta-potentials has been a subject of great interest in recent years \cite{ro1,ro2,ro3,ro4}. In solid state quantum
     computation, finite $\delta$-function potentials arrays are often used to describe an instantaneous interaction between flying qubits and
     statics qubits \cite{bos,cic1,cic2}.
     
     The problem of one-dimensional delta-function potential array has been studied in previus works\cite{ita,ya1,ya2,ya3,ya4}. In reference
     \cite{ita} they use a field-theoretic approach to obtain the transfer matrix in terms of a propagator. In references
     \cite{ya1,ya2,ya3,ya4} the case of two delta-function potentials is studied using a quantum-mechanical approach. We use a similar aproach
     to develop a convenient way to deal with the problem, based in a transfer matrix methodology. 
     
     We suppose a particle incident from the left on a linear array of $N$ delta-function potentials. Each potential reflects and transmits a
     part of the particle wave function, which is taken to be a linear combination of an incoming and an outgoing plane wave in between two
     potentials.
     
     This paper is organized as follows. In section \ref{matrix} the transfer matrix is obtained and some of its interesting properties are
     discussed. In section \ref{regular} results for specific regular arrays and arbitrary $N$ are presented. In section \ref{two} we analyze
     some particular symmetries and invariances of the delta-function potential array for $N = 2$, which has many interesting features. In that
     section we also analyze a condensed matter system: the transmission probability of one electron scattered by two impurities in a GaAs
     nanowire. Finally, section \ref{cr} contains some concluding remarks.
     
\section{The transfer matrix} \label{matrix} 
     
     We study the motion of a particle incident on $N$ delta-function potentials located on a one-dimensional quantum wire along the
     $x$-axis. The Hamiltonian of the system is   
     
     \begin{equation} 
     \label{h}
     \hat{H}=\frac{\hat{p}^{2}}{2m} + \sum_{n = 1}^N J_n \delta(x - x_n),          
     \end{equation} 
     
     \noindent
     where $\hat{p}$ and $m$ are momentum and mass of the particle. The values $x_n$ and $J_n$ are the position and the strength of the $n$-th
     delta-function potential. The sign of $J_n$ can be positive denoting a potential barrier or negative denoting a potential well.
     
     \begin{figure}[bt]
     \centerline{\psfig{file=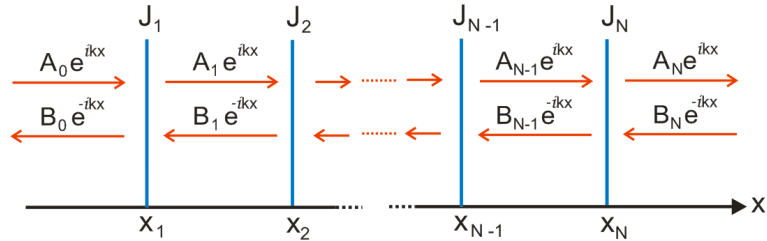 ,width=4.10in}}
     \vspace*{8pt}
     \caption{(Color online) One dimensional quantum wire with $N$ delta potentials of strength $J_n$ located at $x=x_n$,
     $n=1,2,..,N$.}
     \label{Fig1}
     \end{figure} 
     
     Figure \ref{Fig1} shows the array of the $N$ delta-function potentials. In this case an incoming particle from the left will be partially
     reflected and partially transmitted by each delta-function potential. As a result of these reflections and transmissions the
     particle wave function will be described by a combination of an incoming and an outgoing wave in between two potentials. The only
     exception is in the zone after the last ($N$-th) potential. That is, in the region $x > x_N$ there will be only an outgoing wave 
     ($e^{\imath kx}$) since no particle is incident from the right. For each of the $N+1$ zones in which the potentials divide the axis, the
     wave function is taken to be 
     
     \begin{equation}
     \label{fun}
     \psi_n (x) = A_n e^{\imath kx} + B_n e^{-\imath kx} \ \ \ \ \ \ \ \ \ \ \ \ x_{n} \leq x \leq x_{n+1}.          
     \end{equation}
     
     The wave amplitudes $A_n$ and $B_n$ are constant coefficients and $k$ is the wave number for the particle energy 
     $\epsilon = \hbar^2k^2/2m$. This wave function is continuous at every point on the $x$-axis, however the delta potentials produce a
     discontinuity in the derivative of the wave function at the points $x_n$, which is given by
     
     \begin{equation}
     \label{dis}
     \left(\frac{d \psi}{dx}\right)_{x=x_n+}-\left(\frac{d \psi}{dx}\right)_{x=x_n-}=\frac{2mJ_n}{\hbar^2}\psi(x_n).          
     \end{equation} 
     
     \noindent
     For the $n$-th potential $J_n\delta(x-x_n)$ at $x=x_n$, the continuity of the wave function yields the relation
     
     \begin{equation}
     \label{con}
      A_{n-1} e^{\imath kx_n} + B_{n-1} e^{-\imath kx_n} = A_{n} e^{\imath kx_n} + B_{n} e^{-\imath kx_n}.          
     \end{equation}
     
     \noindent
     Defining $E_n \equiv e^{2\imath kx_n}$, Eq. \ref{con} becomes
     
     \begin{equation}
     \label{con2} 
      A_{n} E_n + B_{n} =  A_{n-1} E_n + B_{n-1},          
     \end{equation}
    
     \noindent
     From the discontinuity in the derivative of the wave function at $x=x_n$ we obtain
     
     \begin{multline} 
     \label{dis2}
      \imath k(A_{n}e^{\imath kx_n}  - B_{n} e^{-\imath kx_n}-A_{n-1}e^{\imath kx_n}  + B_{n-1} e^{-\imath kx_n}) \\
      =\frac{2mJ_n}{\hbar^2}(A_{n}e^{\imath kx_n}  + B_{n} e^{-\imath kx_n})
     \end{multline}
     
     \noindent
     or
      
     \begin{equation} 
     \label{dis3}
      (1+\imath \lambda_n)A_n E_n-(1-\imath \lambda_n)B_n = A_{n-1} E_n-B_{n-i},        
     \end{equation}
     
     \noindent
     in terms of the dimensionless parameter $\lambda_n=2mJ_n/\hbar^2k$. Equations (\ref{con2}) and (\ref{dis3}) form a system of linear
     independent equations which can be expressed in matrix form as 
     
     \begin{equation}
     \label{mat}
        \left(\begin{array}{c} A_n \\ B_n \end{array}\right)={\bf M}_n\left(\begin{array}{c} A_{n-1} \\ B_{n-1} \end{array}\right),
     \end{equation}
     
     \noindent
     where
     
     \begin{equation}
     \label{mat2}
          {\bf M}_n =\frac{1}{2}\left( \begin{array}{cc}
                  2-\imath \lambda_n  &  -\imath \lambda_n E_n^*  \\
                 \imath \lambda_n E_n & 2+\imath \lambda_n \end{array} \right).
      \end{equation}

     ${\bf M}_n$ is a transfer matrix which relates the coefficients of the wave function in the $n$-th zone with those in the ($n-1$)-th zone.
     As can be observed ${\bf M_n}$ is a non-unitary matrix with determinant equal to 1, it can be written in terms of the identity matrix
     ${\bf I}$ as
     
     \begin{equation}
     \label{M}
      {\bf M}_i={\bf I}-\frac{\imath\lambda_n}{2}{\bf L}_n    
     \end{equation} 
     
     \noindent
     where 
     
     \begin{equation}
     \label{L}
      {\bf L}_n(E_n) =  \left( \begin{array}{cc}
                  1              &  E_n^*  \\
                  -E_n             & -1              \end{array} \right). 
     \end{equation} 
     
     \noindent
     Equation (\ref{M}) divides ${\bf M}_n$ in a convenient way to separate the interaction. Further ${\bf L}_n$ only depends of the
     dimensionless parameter $kx_n$. It is remarkable that this dependence is only in the off-diagonal elements of ${\bf L}_n$, showing that
     the effect of the incoming wave over the outcoming one differs in only a phase. 
     
     Now we can use Eq. (\ref{mat}) to relate the wave function coefficients of the outgoing wave function in the zone $N$, to the coefficients
     of the incident wave function to obtain
          
     \begin{equation}
     \label{mm}
          \left(\begin{array}{c} A_N \\ B_N \end{array}\right) \equiv {\bf \mathcal{M}}(N)
          \left(\begin{array}{c} A_{0} \\ B_{0} \end{array}\right),
     \end{equation}  
    
    \noindent    
    with ${\bf \mathcal{M}}(N)\equiv{\bf M}_N{\bf M}_{N-1}...{\bf M}_2{\bf M}_1$. This formalism allow us to analyze the problem of a
    particle incident from the left on an array of $N$ delta-function potentials. After the last potential, at $x=x_N$, there is only an
    outgoing wave to the right, since we assume there is no particle (wave) incident from the right of the $N$-th potential, that is $B_N =
    0$. Since $\det({\bf M}_n)=1$ for all $n$ and $B_N = 0$, Eq.(\ref{mm}) gives    
    
    \begin{equation}
    \label{an}
         A_N= \frac{A_0}{({\bf \mathcal{M}}(N))_{22}}.
     \end{equation} 
    
    \noindent
    The coefficient $A_N$ is the transmission probability amplitude which means that the the probability of transmition is  $T
    =|A_N|^2/|A_0|^2$. The incoming and outcoming flux have to be equal in each zone this means that
    $|A_n|^2-|B_{n}|^2=|A_{n-1}|^2-|B_{n-1}|^2$ for $n=1,2,...,N$. 
    
    We note here some interesting properties of the ${\bf L}_n$ which will be useful in the next sections. The anticommutator of the 
    ${\bf L}$-matrices is
     
     \begin{equation}
     \label{antic}
         {\bf L}_n{\bf L}_m+{\bf L}_m{\bf L}_n = (2-E_nE_m^*-E_n^*E_m){\bf I}.
     \end{equation} 
     
     \noindent
     Taking $E_n = E_m$, we note that ${\bf L}_n^2 = 0$ which implies that ${\bf M}_n^{-1}={\bf I}+\frac{\imath\lambda_n}{2}{\bf L}_n$. 
     Also, if $E_n = -E_m$ then ${\bf L}_n{\bf L}_m+{\bf L}_m{\bf L}_n= 4{\bf I}$. Now, defining $\phi_{mn} = k(x_m - x_n)$ Eq. (\ref{antic})
     can be expressed as
     
     \begin{equation}
     \label{antic2}
         {\bf L}_n{\bf L}_m+{\bf L}_m{\bf L}_n = 4\sin^2(\phi_{mn}){\bf I}.
     \end{equation}
    
    \noindent
    Equation (\ref{antic2}) is useful to simplify the multiple products of $\bf{L}$'s present in the transfer matrix. From Eq. (\ref{mm}), the
    general structure of the full transfer matrix ${\bf \mathcal{M}}(N)$ in terms of $\lambda_n{\bf L}_n$ ($n=1,2,..,N$) is clear 
    
    \begin{multline}
    \label{mmex}
        {\bf \mathcal{M}}(N)= {\bf I}-\frac{\imath}{2}\sum_{n_1=1}^N\lambda_{n_1}{\bf L}_{n_1}  
         -\frac{1}{4}\sum_{\substack{n_1,n_2=1\\n_1>n_2}}^N \lambda_{n_1}{\bf L}_{n_1}\lambda_{n_2}{\bf L}_{n_2}+\\
        ...+\left(-\frac{\imath}{2}\right)^m\sum_{\substack{n_1,n_2,..,n_m=1\\n_1>n_2>...>n_m}}^N 
        \lambda_{n_1}{\bf L}_{n_1}\lambda_{n_2}{\bf L}_{n_2}...\lambda_{n_m}{\bf L}_{n_m}+ \\
        ...+\left(-\frac{\imath}{2}\right)^N\lambda_{n_1}{\bf L}_{n_1}\lambda_{n_2}{\bf L}_{n_2}...\lambda_{n_N}{\bf L}_{n_N}.
    \end{multline}  
        
    \noindent
    If all $\lambda_n$ are zero, that is no potentials are present, then obviously ${\bf \mathcal{M}}={\bf I}$. Equation (\ref{mmex})
    simplifies enormously for special cases. For example, if all $\lambda_n = \lambda$ for $n=1,2,..,N$, ${\bf \mathcal{M}}(N)$ is polynomial
    of order $N$ in the strength $\lambda$. The information of the location of the potentials ($x_n$) appears in ${\bf L}_n$ through
    $E_n=e^{2\imath kx_n}$. Moreover, since products of ${\bf L}_n$ appears in Eq. (\ref{mmex}) it is through $E_nE_m^*$. Consequently, for
    specific regular arrays one can use Eq. (\ref{antic2}) to simplify ${\bf \mathcal{M}}(N)$ when there are specific relations between the
    $E_n$'s.

\section{Regular arrays} \label{regular}
 
    Consider a regular array in which $x_n-x_{n-1}\equiv d$ for $n=2,3,...,N$. The transfer matrix will be different for particles for 
    different $k$. For a regular array the dimensionless quantity $kd$ plays the crucial role. We study a couple of specific
    examples for choice of $kd$.
    
    {\bf Case A}. The simplest case is when $kd = \pi$. This situation is known as resonant condition (RC) and is widely used to describe
    spin scattering \cite{cic1,cic2,com}. In this case $E_nE_{n+1}^*=1$, consequently all $E_n=E$ and ${\bf L}_n=L$ for $n=1,2,3,...,N$.
    Since ${\bf L}^2=0$, Eq. \ref{mmex} reduces to simply
    
    \begin{equation}
    \label{mmA}
       {\bf \mathcal{M}}(N)= {\bf I}-\frac{\imath}{2}{\bf L}\sum_{n=1}^N\lambda_n.
    \end{equation} 
    
    In this case ${\bf \mathcal{M}}$ is linear and symmetric in the $\lambda_n$'s. The transmission probability (see Eq. (\ref{an})) is
    
     \begin{equation}
     \label{TA}
       T = \frac{1}{|1+\frac{\imath}{2}\sum_{n=1}^N\lambda_n|^2}=\frac{4}{4+(\sum_{n=1}^N\lambda_n)^2}.
     \end{equation} 
     
     \noindent
     Moreover, if the potentials are both attractive (negative $\lambda$) or repulsive (positive $\lambda$) this can profoundly affect $T$.
     In fact $T=1$ if the sum $\sum_{n=1}^N\lambda_n=0$, the individual values of the $\lambda_n$ do not matter  as long as the sum is zero.
     
     {\bf Case B}. Consider the case when $kd=\pi/2$. In this case $E_nE_{n+1}^*=-1$. Consequently, $E_{2n+1}=E_1$ and $E_{2n}=-E_1$ with
     $n=1,2,..$ . Correspondingly ${\bf L}_{2n+1}={\bf L}_1$ and ${\bf L}_{2n}={\bf L}_1^{\dag}$. From Eq. (\ref{antic2}) 
     ${\bf L}_1{\bf L}_1^{\dag}+{\bf L}_1^{\dag}{\bf L}_1 = 4{\bf I}$ and ${\bf L}_1{\bf L}_1={\bf L}^{\dag}_1{\bf L}^{\dag}_1=0$, it
     can be shown that ${\bf \mathcal{M}}$ only contains linear terms in ${\bf L}_1{\bf L}_1^{\dag}$ and ${\bf L}_1^{\dag}{\bf L}_1$ for any
     given $N$. For $N=4$ $E_1=E_3=E$, $E_2=E_4=-E$ and ${\bf L}_1={\bf L}_3={\bf L}$, ${\bf L}_2={\bf L}_4={\bf L}^{\dag}$. In this case 
     ${\bf \mathcal{M}}(4)$ reduces to 
     
     \begin{multline}
     \label{mm4}
      {\bf \mathcal{M}}(4) = (1-\lambda_3\lambda_2){\bf I}-\frac{\imath}{2} 
      ((\lambda_3+\lambda_1){\bf L}+(\lambda_4+\lambda_2){\bf L^\dag})
      - \frac{1}{4}(\lambda_4\lambda_3+\lambda_4\lambda_1
      +\lambda_2\lambda_1-\lambda_3\lambda_2){\bf L^\dag L}\\
     +\frac{\imath}{2}(\lambda_3\lambda_2\lambda_1{\bf L}+\lambda_4\lambda_3\lambda_2{\bf L^\dag})
     +\frac{1}{4}\lambda_4\lambda_3\lambda_2\lambda_1
      {\bf L^\dag L}. 
     \end{multline}
     
     \noindent 
     Consequently
     
     \begin{multline}
     \label{mm42}
      ({\bf \mathcal{M}}(4))_{22}= (1-\lambda_3\lambda_2) -\frac{\imath}{2} 
      (\lambda_3+\lambda_1+\lambda_4+\lambda_2) 
      - \frac{1}{2}(\lambda_4\lambda_3+\lambda_4\lambda_1+ \lambda_2\lambda_1-\lambda_3\lambda_2)\\
     +\frac{\imath}{2}(\lambda_3\lambda_2\lambda_1+\lambda_4\lambda_3\lambda_2) +\frac{1}{2}(\lambda_4\lambda_3\lambda_2\lambda_1). 
     \end{multline}
        
     We note that the case B can be reduced to case A by changing the wave number of the particle $k \rightarrow 2k$. Below we discuss the
     specific example of $N=2$ in detail.

\section{Two delta-potential system} \label{two}

    We now focus on a $N=2$ array. Although it is the simplest array, it presents interesting behavior like resonant tunneling. In this case
    
    \begin{equation}
    \label{mm2}
       {\bf \mathcal{M}}(2)= {\bf I}-\frac{\imath\lambda_2}{2}{\bf L}_2-\frac{\imath\lambda_1}{2}{\bf L}_1
       -\frac{\lambda_2\lambda_1}{4}{\bf L}_2{\bf L}_1,
     \end{equation} 
    
    \noindent
    and the transmission probability when the incoming flux has $A_0=1$, (see Eq. (\ref{an}) ) is
    
    \begin{equation}
    \label{T2}
       T = \frac{1}{|({\bf \mathcal{M}}(2))_{22}|^2}
       =\frac{16}{|(2+\imath\lambda_2)(2+\imath\lambda_1)+\lambda_2\lambda_1E_2E_1^*|^2}.
     \end{equation} 
     
     Equation (\ref{T2}) shows that $T$ depends only on the distance between potentials ($x_2-x_1$). Further, $T$ is symmetric under
     interchange of $\lambda_1$ and $\lambda_2$ independently of the location of the potentials. The multiscattering present in the middle of
     the two potential can create a positive interference between the waves, and therefore, this can produce  a resonant tunneling.
     Nevertheless, an extra phase is added to the wave function when it interacts with every potential increasing the complexity of the
     problem. To find the conditions when resonant tunneling can happen, we calculate the derivative of $|({\bf \mathcal{M}}(2))_{22}|^2$ with
     respect to $\phi_{21} = k(x_2 - x_1)$. The vanishing of the derivative gives the condition
          
     \begin{equation}
     \label{pp}
       \tan 2\phi_{21}=\frac{2(\lambda_1+\lambda_2)}{4-\lambda_2\lambda_1}.
    \end{equation}
    
    \begin{figure}[bt]
     \centerline{\psfig{file=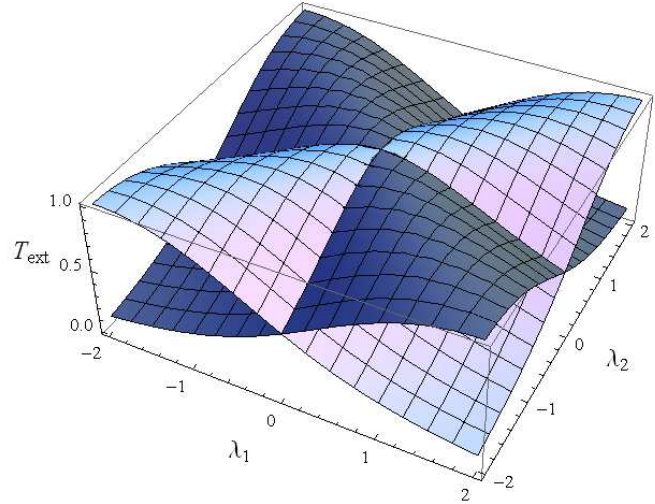,width=3.50in}}
     \vspace*{8pt}
     \caption{\label{Fig2}
      (Color online) $T_{ext}$ as a function of $\lambda_1$ and $\lambda_2$ varying between $-2$ and $2$. The white surface represents
     the case where $n$ is an odd integer, and the gray surface represents the case where $n$ is an even integer. Notice that for odd (even)
      $n$ and $\lambda_1=\lambda_2$ ($\lambda_1=-\lambda_2$) we get perfect transmission ($T=1$).}
     \end{figure} 
    
    \noindent
    Note that Eq. (\ref{pp}) is satisfied by $2\phi_{21}+n\pi$, for integer $n$. This implies interesting behavior of $T$ when the distance
    between the potentials is varied keeping the strengths $\lambda_1$ and $\lambda_2$ fixed. Keeping the first potential fixed at $x=x_1$,
    let the position of the second potential be at $x_2(n)= x_2+n\pi / 2k$, $n=0,\pm 1,\pm 2,...$. In this case, using Eq. \ref{pp}, the
    extremal values of the transmission probability ($T_{ext}$) are given by
    
    \begin{equation}
    \label{Tex}
    T_{ext}= \\
    \frac{8}{4+\Lambda_1\Lambda_2+(-1)^n\lambda_1\lambda_2\sqrt{(2+\Lambda_1)(2+\Lambda_2)}},
    \end{equation}

    \noindent
    with $\Lambda_n = \lambda_n^2+2$. If $n$ is an odd (even) integer, the value of $T$ is maximum (minimum) when both $\lambda_1$ and
    $\lambda_2$ have the same sign. If $\lambda_1$ and $\lambda_2$ have opposite sign then $T$ is maximum (minimum) for $n$ even (odd)
    integer. The crucial point is the sign of the last term in the denominator of Eq. (\ref{Tex}). The behavior of $T_{ext}$ as a function of
    $\lambda_1$ and $\lambda_2$ is displayed graphically in Fig. \ref{Fig2} for even and odd integer. When the magnitude of the strength of
    both potentials is equal ($\lambda_1=\lambda=\pm \lambda_2$), Eq. (\ref{Tex}) becomes to
    
    \begin{equation}
    \label{Tex2}
    T_{ext}=\frac{8}{8 + \lambda^2(4 +\lambda^2)(1 \pm (-1)^n)}.
    \end{equation} 
    
    \noindent
    For odd (even) $n$ and $\lambda_1=\lambda_2$ ($\lambda_1=-\lambda_2$) we get perfect transmission ($T=1$). Note that when we have an even 
    $n$ and $\lambda_1=-\lambda_2$, then $\phi_{21}=m\pi$, and we obtain the case A described in section \ref{regular}.

    The perfect transmition ($T=1$) is present in the system where $\lambda_1 = \pm \lambda_2$ due to the symmetry of the
    system. Notice that $T=1$ with $\lambda_1 = -\lambda_2$ and $\phi_{12}=n\pi$ independently of the values of the $\lambda$'s which implies
    that the extra phase added to the wave function in the scattering is equal to zero. In all other situations the extra phase depends on the
    values of the $\lambda$'s.
    
     \begin{figure}[bt]
     \centerline{\psfig{file=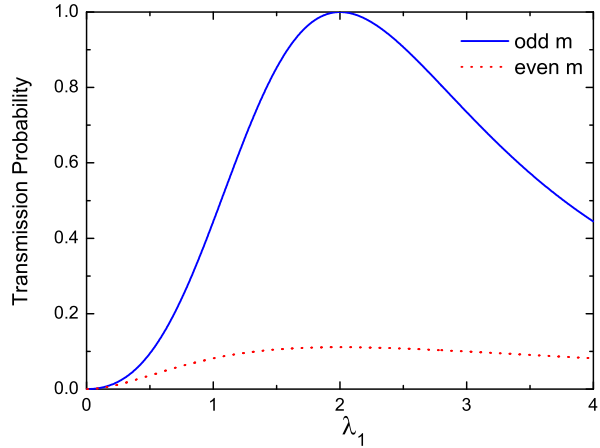,width=3.65in}}
     \vspace*{8pt}
     \caption{\label{Fig3}
     (Color online) Transmission probability $T$ as a function of $\lambda_1$ when $\lambda_2= 4/\lambda_1$. The blue (solid) line
     shows the case when  $m$ is an odd integer and $T$ is a maximum. The red (dotted) line shows the case when $m$ is even and $T$ is the
     minimum.}
     \end{figure}  
    
    For the particular case when $\lambda_1\lambda_2=4$, Eq. (\ref{pp}) requires that $\cos 2\phi_{21}=0$. In fact for
    $2\phi_{21}=(2m+1)\pi/2$, ($m=0,1,2...$), $E_2E_1^* = \exp[\imath(2m+1)\pi/2]$ and this is equal to $(-1)^m\imath$. In this case,
    Eq. (\ref{T2}) reduces to
    
    \begin{equation}
    \label{Tex3}
    T=\frac{4}{12+\lambda_1^2+\lambda_2^2+4(-1)^m(\lambda_1+\lambda_2)}.
    \end{equation} 
    
    \noindent
    The condition $\lambda_1\lambda_2=4$ requires both the potential strength parameters to have the same sign, even so one can obtain $T=1$.
    In Fig. \ref{Fig3} $T$ is plotted as a function of $\lambda_1$ when $\lambda_2= 4/\lambda_1$. One obtains $T=1$ for $\lambda_1=\lambda_2=
    2$ when $m$ is an odd integer, and also for $\lambda_1=\lambda_2=-2$ when $m$ is an even integer. This is really interesting because the
    potentials are both repulsive or attractive. The separation of the potentials is crucial in producing resonant tunneling. 
        
\begin{figure}[bt]
     \centerline{\psfig{file=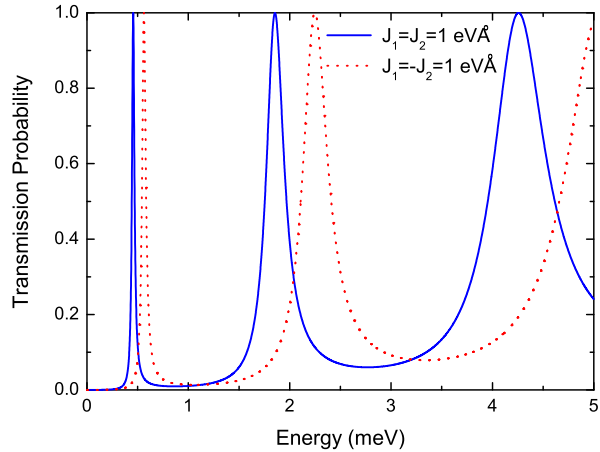,width=3.65in}}
     \vspace*{8pt}
     \caption{\label{Fig4}
     (Color online) Transmission probability $T$ as a function of the electron energy in a two-potential array, when the effective
     mass of electron is 0.067 times the electron mass, the distance between potentials is $x_1-x_2=100$ nm and $\lambda_1=\lambda_2=1$ 
     eV \AA (solid blue line), or $\lambda_1=-\lambda_2=1$ eV \AA (dotted red line).}
     \end{figure}

    To set these ideas in a condensed matter system, we suppose that the two $\delta$-function potentials represent two impurities fixed
    in a GaAs quantum wire, and we study the transmission of a ballistic electron, although the formulation in this paper can apply to any
    particle which interacts with delta-function-like potentials. In this situation, the effective mass is 0.067 times the electron mass,
    and the strengths of interaction are $J_1=J_2 \simeq 1$ eV \AA as reported for magnetic impurities \cite{cic1,mei}. Figure \ref{Fig4} 
    shows the transmission probability as a function of the electron energy if the distance between impurities is $x_1-x_2=1000$ \AA.

    The curves of $T$ as a function of the electron energy in the cases when $\lambda_1=\lambda_2=1$ eV \AA and $\lambda_1=-\lambda_2=1$ 
    eV \AA are very similar, but the influence of the extra phase present in the first case ($\lambda_1=\lambda_2$), produces a
    displacement in the values of the resonant energy. Notice that the peaks denoting the energies where there is perfect transmission
    (resonant energies) are more and more wide as the energy is increased, suggesting an already expected asymptotic behavior of $T$ for
    perfect transmission for large values of energy.

\section{Conclusions} \label{cr}

    In this paper we studied the problem of one-dimensional quantum wire along which a moving particle interacts with a linear array of $N$
    delta-function potentials using a transfer matrix method. We showed that the transfer matrix (${\bf M}_n$) for the $n$-th potential,
    can be expressed in terms of the unit matrix and a matrix ${\bf L}_n$, which has interesting properties and useful anticommutation
    relations. The properties of the ${\bf L}_n$ matrix were used to calculate the transmission probability in the case of regular arrays. 
    
    The case of the simplest array, namely $N=2$, was considered in detail in section \ref{two}. This unexpectedly showed a variety of cases in
    which resonant conditions are possible. One obtained $T=1$ even when $\lambda_1$ and $\lambda_2$ have the same sign. As shown above this
    is due to the interplay between the values of $\lambda_1$, $\lambda_2$, the separation ($x_2-x_1$) between potentials and the wave number
    $k$ of the scattered particle. We also show that $T=1$ is only possible when $\lambda_1 = \pm \lambda_2$ and for certain values of
    $\phi_{21}=k(x_2-x_1)$.The methodology developed here could be useful for the study of different potential arrays. We have obtained the
    transmitted wave function from all the information that defines the potential array (positions and strengths of the potentials). It could
    be also interesting to study the inverse situation: How much information about the potential array could be obtained from a scattering
    experiment? This could be useful in the implementation of a scattering-based quantum information system. 

\section*{Acknowledgements}

     One of the autors, G. C. gratefully acknowledges financial support from CONACYT (Mexico). This work was supported by CONACYT under grant
     No. 83604.


\begin{thebibliography}{0}
\bibitem{com} In this paper, differently from Refs. ~\refcite{cic1,cic2} the term resonant conditions refers to the conditions under which maximum
   transmission probability is obtained.
\end{thebibliography}
\end{document}